# A unified theory of cataclysmic variables from self-consistent numerical simulations

Yael Hillman[1], Michael M. Shara[2], Dina Prialnik[3] and Attay Kovetz[4]

[1]Department of Physics, Ariel University, Ariel POB 3, 40700, Israel

[2]Department of Astrophysics, American Museum of Natural History, CPW & 79th street, New York, NY 10024, USA

[3]Department of Geosciences, Tel Aviv University, Ramat Aviv, Tel Aviv 69978, Israel

[4]School of Physics and Astronomy, Faculty of Exact Sciences, Tel-Aviv University, Tel Aviv, Israel

**The hydrogen-rich envelopes accreted by white dwarf stars from their red dwarf companions[1] lead to thermonuclear runaways[2,3] observed as classical nova eruptions[4,5] peaking at $10^{5-6}$ solar luminosities. Virtually all nova progenitors are novalike binaries exhibiting high rates of mass transfer to their white dwarfs before and after an eruption[6]. It is a puzzle that binaries indistinguishable from novalikes, but with much lower mass transfer rates[7], and resulting dwarf nova outbursts[8], co-exist at the same orbital periods. Nova shells surrounding several dwarf novae[9-12] demonstrate that at least some novae become dwarf novae between successive nova eruptions[13-14], though the mechanisms and timescales governing mass transfer rate variations are poorly understood. Here we report simulations of the multi-Gyr**

**evolution of novae which self-consistently model every eruption's thermonuclear runaway, mass and angular momentum losses, feedback due to irradiation and variable mass transfer, and orbital size and period changes. The simulations reproduce the observed wide range of mass transfer rates at a given orbital period, with large and cyclic changes in white dwarf - red dwarf binaries emerging on kyr – Myr timescales. They also demonstrate that deep hibernation – complete stoppage of mass transfer for long periods - occurs only in short-period binaries; that initially very different binaries converge to become nearly identical systems; that while almost all pre-novae should be novalike binaries, dwarf novae should also occasionally give rise to novae; and that the masses of white dwarfs decrease only slightly while their red dwarf companions are consumed.**

Mass and angular momentum transfer from the red dwarf to the white dwarf component of cataclysmic variable binary stars drive a rich variety of variability as well as the evolution of these systems[7]. Winds from the red dwarf and from the white dwarf's accretion disk[15], and shells ejected during nova events[16] carry away much of the mass and angular momentum of cataclysmic variables over their multi-Gyr lifetimes. Most simulations to date of cataclysmic variable evolution over Gyr have assumed constant mass transfer rates between nova outbursts[5] and monotonically decreasing rates over Gyr of evolution[7]. However, nova eruptions subject red dwarfs to irradiation exceeding their own luminosity by factors of hundreds for decades, which should swell red dwarfs and increase mass transfer rates[17,18]. Mass ejected during nova eruptions should increase the separation between a cataclysmic variable's red dwarf and white dwarf[14], decreasing the mass transfer rate. The timescales for these competing processes to significantly change the appearances of cataclysmic variables are a century or longer, so existing observational baselines are too short to determine the relative importance of these feedback effects in driving these stars' evolution.

This work takes the evolution of cataclysmic variables one step further by implementing feedback between the red dwarf and white dwarf (see Methods) due to the effects noted above, over the entire, multi-Gyr evolution of cataclysmic variables. The key aim is to quantify mass transfer rates ($\dot{M}$) in these binaries throughout their lifetimes. We followed the self-consistent, continuous evolution of four cataclysmic variable models from their births until the donor mass was eroded down to the hydrogen-burning limit at $0.09 M_\odot$. The red dwarfs were assumed to be unevolved when mass transfer began; evolved secondaries are known[19] and will be studied elsewhere. Every one of the thousands of nova eruptions that each binary underwent was simulated in great detail (see Methods).

The initial models are summarized in Table 1. The white dwarf and red dwarf masses cover a significant region of cataclysmic variable parameter space and allow a systematic comparison between the evolution of different initial stellar mass combinations. The onset of red dwarf to white dwarf mass transfer, immediately after a nova eruption, is the beginning of a nova cycle, while each cycle ends with a nova eruption yielding mass loss and luminosity decline. Our simulations show that the inclusion of feedback causes $\dot{M}$ to vary by orders of magnitude during the accretion phase of each nova cycle, and that it experiences very large secular changes over thousands of cycles spanning several Gyr.

Figure 1 presents the $\dot{M}$ histories of nine nova cycles at different epochs during the multi-Gyr evolution of each model, illustrating the long and short-term trends of $\dot{M}$ variation. During, and soon after a nova eruption the red dwarf is strongly irradiated by the white dwarf, enhancing Roche Lobe overflow and hence $\dot{M}$. $\dot{M}$ then decreases over a few hundred to thousands of years, as the effect of the cooling white dwarf's irradiation diminishes. Eventually magnetic braking and gravitational radiation reverse the decreasing $\dot{M}$ trend by decreasing the binary separation. Thus, irradiation-driven $\dot{M}$ begins high at the start of each cycle, decreases relatively quickly, and then much more slowly and monotonically increases until the total amount of accreted mass is sufficient to trigger the next nova eruption. $\dot{M}$ at the beginning of each nova cycle decreases monotonically throughout a binary's evolution. When the temporary effect of irradiation subsides after a nova eruption, $\dot{M}$ is always lower than it was at the same stage of the previous cycle. Since a longer interval is needed in each successive cycle to reach the critical mass for a nova outburst, the time between consecutive nova eruptions increases as the evolution progresses.

    The mass transfer rate at the beginning of a cycle is determined by the binary component masses and separation at that time. This is seen in Figure 1 by comparing two models when the red dwarf and white dwarf masses are the same, which occurs at different times in the two cases. For example, cycle 850 of model 1 and cycle 3500 of model 2 (both are purple curves), both with $M_{WD} \simeq 0.7 M_\odot$ and $M_{RD} \simeq 0.45 M_\odot$, each begin with $\dot{M} \sim 10^{-11} M_\odot yr^{-1}$, although their evolutionary times (ages) are very different. The $\dot{M}$ histories for both these cycles are very similar, with $\dot{M}$ decreasing by almost an order of magnitude from that at the start of the cycle. $\dot{M}$ then increases to a final value $\sim 10^{-8} M_\odot yr^{-1}$ in $\lesssim 10^5$ years. Such systems would appear as novalike variables for the part of the nova cycle when $\dot{M} \sim 10^{-8} M_\odot yr^{-1}$, and dwarf novae when $\dot{M}$ is ~10-1000X

smaller. That the same $\dot{M}$ behavior occurs in similar-mass binary systems with initially very different component masses means that the history of a binary system is not important in determining the outcome or aftermath of a nova outburst.

Figure 2 (top panel) shows that the white dwarf masses decrease monotonically, but by only a few percent throughout the evolution of the four cataclysmic variable models. Because of this, cataclysmic variables with similar initial white dwarf masses but different initial red dwarf masses converge to become nearly identical binaries[20]. This occurs as the mass of the initially more massive red dwarf decreases and approaches that of the initially less massive red dwarf. The mass transfer rates are always far too low to lead to nonexplosive hydrogen or helium burning, essential to grow any white dwarf to the Chandrasekhar limit and a type Ia supernova[21]. Such binaries are likely to require secondaries which have evolved beyond the main sequence before coming into contact with their Roche lobes.

In contrast, models with different white dwarf masses, but identical initial red dwarf masses experience entirely different histories. Comparing model 1 ($M_{WD} = 0.7 M_\odot$, $M_{RD} = 0.45\,M_\odot$) with model 3 ($M_{WD} = 1.0 M_\odot$, $M_{RD} = 0.45 M_\odot$) in Figure 1, we see that the number of cycles required to erode the same initial red dwarf mass to $0.09 M_\odot$ is more than three times larger for the $1.0 M_\odot$ cataclysmic variable. This is because the accreted mass required for triggering a nova outburst is lower for more massive white dwarfs, hence more cycles are needed in order to remove the same amount of mass from the donor. In all four models the time until the red dwarf mass falls below the hydrogen-burning limit is a few times $10^9$ yr.

After each nova outburst $\dot{M}$ decreases significantly, becoming less than $\sim 10^{-12-13} M_\odot yr^{-1}$ for part of each cycle when the red dwarf masses are $\lesssim 0.3 M_\odot$. Such binaries, with undetectably small $\dot{M}$, would be classified as "detached". They could easily be mistaken for "pre-cataclysmic variables", white dwarf - red dwarf binaries which have not yet come into contact. White dwarf – red dwarf binaries with undetectably low $\dot{M}$ between nova eruptions are actually deeply "hibernating" cataclysmic variables[14], in temporary states of detachment that Figure 1 shows can last $10^6$ yr or more. Any such binaries caught within ~1-2000 yr after a nova outburst could be pinpointed as deeply hibernating cataclysmic variables if still surrounded by old nova shells.

Since $\dot{M}$ at the beginning of each cycle decreases monotonically, nova recurrence times lengthen from $\sim 10^4$ to almost $10^7$ yr over the lifetimes of cataclysmic variables (see middle panel of Figure 2). The most rapid decreases in binary system mass and period (bottom panel of Figure 2), and concomitant increases in nova recurrence times occur quickly, in just the first ~300 Myr of a cataclysmic binary's life. Binary evolution then slows a hundredfold, as angular momentum losses due to magnetic braking cease.

Because nova systems are only recognized when they erupt, the sharp increase in recurrence time with increasing binary age and decreasing red dwarf mass leads to strong observational biases, quantified in Figures 3 and 4. During the past ~200 years, in which novae have been recognized and recorded, those with the shortest recurrence times are ~100 times more likely to have erupted than those with the longest recurrence times. This explains why most known novae are measured to have orbital periods longer than 3 hours (see Figure 3 and ref. 22), even though the lifetimes of cataclysmic variables at long periods are just a few percent of cataclysmic variable lifetimes (see Figure 4). This observational bias also explains why observed nova progenitors exhibit high and equal rates of mass transfer to their white dwarfs before and after an eruption[6], despite the simulations' prediction (see Figure 1) that novae with low mass companions quickly become much fainter after a nova eruption than immediately before it. Such cataclysmic variables, with red dwarf masses of 0.3 $M_\odot$ or less, have very small $\dot{M}$ averaged over entire cycles, and thus they erupt very infrequently and are rarely detected. These rare novae erupt on white dwarfs accreting at rates as low as $\sim 10^{-10} M_\odot yr^{-1}$ (see Figure 1), and they will exhibit dwarf nova eruptions in the years before a nova event[23].

The self-consistent, feedback-dominated evolution of interacting white dwarf - red dwarf binaries that we have described here naturally explains the existence of a wide range of $\dot{M}$ at all orbital periods. The simulations also demonstrate that white dwarf masses barely change while their red dwarf donors are almost entirely depleted. Binaries with similar initial white dwarf masses, but very different red dwarf masses evolve to become nearly identical. In addition, the models predict that cataclysmic variables with orbital periods longer than ~3 hr undergo only moderate $\dot{M}$ variations and hence *mild* hibernation, appearing first as novalike binaries, then as dwarf novae, and again as novalike variables between nova eruptions. Their red dwarfs never lose contact with their Roche lobes, so that mass transfer in these longer period systems never ceases. Magnetic braking and gravitational radiation inexorably force the red dwarfs closer to their white dwarf companions, so that they increasingly overfill their Roche lobes and their $\dot{M}$ values rise until they again appear as novalike binaries before the next nova eruption. Most of the mass needed to achieve a thermonuclear runaway on a white dwarf is accreted during the second high $\dot{M}$ phase, explaining why novalike binaries are the immediate progenitors of most novae, even in short period cataclysmic variables.

The large majority of cataclysmic variables are shorter period systems, erupting only rarely as novae, and occasionally exhibiting dwarf nova eruptions beforehand. The simulations presented here predict that *only* these short period systems undergo very large decreases in $\dot{M}$, and hence *deep* hibernation, appearing as detached binaries for much of the ~ millions of years between their nova eruptions. This distinction between long and short period cataclysmic binaries[24,25] is an essential component of the unified theory of cataclysmic binaries presented here that emerges when numerical modeling includes the self-consistent feedback that dominates these systems' evolution.

**References**

1. Kraft, R. Binary Stars Among Cataclysmic Variables. III. Ten Old Novae. *The Astrophysical Journal* **139**, 457–475 (1964).
2. Starrfield, S., Truran, J.W., Sparks, W. M. & Kutter, G. CNO Abundances and Hydrodynamic Models of the Nova Outburst. *The Astrophysical Journal* **176**, 169–176 (1972).
3. Prialnik, D., Shara, M. & Shaviv, G. The Evolution of a Slow Nova Model with a Z = 0.03 Envelope from Pre-explosion to Extinction. *A&A* **62**, 339–348 (1978).
4. Warner, B. Cataclysmic Variable Stars. *Cambridge Astrophysics Series* **28** (1995).

5. Yaron, O., Prialnik, D., Shara, M. M. & Kovetz, A. An Extended Grid of Nova Models. II. The Parameter Space of Nova Outbursts. *The Astrophysical Journal* **623**, 398–410 (2005).

6. Collazzi, A. C. et al. The Behavior of Novae Light Curves Before Eruption. *The Astronomical Journal* **138**, 1846–1873 (2009).

7. Knigge, C., Baraffe, I. & Patterson, J. The Evolution of Cataclysmic Variables as Revealed by Their Donor Stars. *The Astrophysical Journal Supplement Series* **194**, 28-76 (2011).

8. Dubus, G., Otulakowska-Hypka, M. & Lasota, J.-P. Testing the disk instability model of cataclysmic variables. *A&A* **617,** A26.

9. Shara, M. M. et al. An ancient nova shell around the dwarf nova Z Camelopardalis. *Nature* **446**, 159–162 (2007).

10. Shara, M. M. et al. AT Cnc: A Second Dwarf Nova with a Classical Nova Shell. *The Astrophysical Journal* **758**, 121-126 (2012).

11. Miszalski, B. et al. Discovery of an eclipsing dwarf nova in the ancient nova shell Te 11. *Mon. Not. Roy. Astron. Soc.* **456**, 633-640 (2016).

12. Shara, M. M. et al. Proper-motion age dating of the progeny of Nova Scorpii AD 1437. *Nature* **548**, 558–560 (2017).

13. Vogt, N. The structure and outburst mechanisms of dwarf novae and their evolutionary status among cataclysmic variables. *MitAG* **57**, 79- (1982).

14. Shara, M., Livio, M., Moffat, A. & Orio, M. Do Novae Hibernate During Most of the Millennia Between Eruptions? Links Between Dwarf and Classical Novae, and Implications for the Space Densities and Evolution of Cataclysmic Binaries. *The Astrophysical Journal* **311**, 163–171 (1986).

15. Drew, J. Inclination and orbital-phase-dependent resonance line-profile calculations applied to cataclysmic variable winds. *Mon. Not. Roy. Astron. Soc.* **224**, 595-632 (1987).

16. Gill, C.D. and O'Brien, T. J. Hubble Space Telescope imaging and ground-based spectroscopy of old nova shells - I. FH Ser, V533 Her, BT Mon, DK Lac and V476 Cyg. *Mon. Not. Roy. Astron. Soc.* **314**, 175-182 (2000).

17. Kovetz, A., Prialnik, D. and Shara, M. M. What Does an Erupting Nova Do to its Red Dwarf Companion? *The Astrophysical Journal* **325**, 828-836 (1988).

18. Ritter, H., Zhang, Z. -Y. and Kolb, U. Irradiation and mass transfer in low-mass compact binaries. *A&A* **360**, 969-990 (2000).

19. Baraffe, I. and Kolb, U. On the late spectral types of cataclysmic variable secondaries. *Mon. Not. Roy. Astron. Soc.* **318**, 354-360 (2000).

20. Stehle, R., Ritter, H., and Kolb, U. An Analytic Approach to the secular evolution of cataclysmic variables. *Mon. Not. Roy. Astron. Soc.* **279**, 581-590 (1996).

21. Hillman, Y., Prialnik, D., Kovetz, A. and Shara, M. Growing White Dwarfs to the Chandrasekhar Limit: The Parameter Space of the Single Degenerate SNIa Channel. *The Astrophysical Journal* **819**, 168-178 (2016).

22. Patterson, J. *et al.* BK Lyncis: the oldest old nova and a Bellwether for cataclysmic variable evolution. *Mon. Not. Roy. Astron. Soc.* **434**, 1902-1919 (2013).

23. Mroz, P., Udalski, A., Pietrukowicz, P., Szymanski, M.K., Soszynski, I. et al. *Nature* 537, 649-651 (2016).

24. Townsley, D. and Bildsten, L. Classical Novae as a probe of the cataclysmic variable population. *The Astrophysical Journal* **628**, 395-400 (2005).

25. Livio, M. and Shara, M. M. Binary system parameters and the hibernation models of cataclysmic variables. *The Astrophysical Journal* **319**, 819-826 (1987).

**Acknowledgements**

We thank the dozens of observers who have worked diligently, over the past three decades, to test predictions of the hibernation scenario of cataclysmic variables. We also thank Claus Tappert, Linda Schmidtobreick, Brad Schaefer and Christian Knigge for valuable constructive criticisms of an earlier draft of this paper.

**Author Contributions**

All authors shared in formulating the ideas underlying the simulations, the computer algorithms and the writing of this paper. YH carried out the simulations and the data mining that produced the figures.

**Author Information**

Reprints and permission information is available at www.nature.com/reprints.

**Competing Interests**

The authors declare that they have no competing financial interests.

**Corresponding Authors**

YH and MS are the corresponding authors. Correspondence and requests for materials should be addressed to yaelhi@ariel.ac.il or mshara@amnh.org .

**Data Availability**

All data pertaining to each simulation is available upon reasonable request from YH.

**Table**

| Model | $M_{WD}[M_\odot]$ | $M_{RD}[M_\odot]$ |
|---|---|---|
| 1 | 0.7 | 0.45 |
| 2 | 0.7 | 0.7 |
| 3 | 1.0 | 0.45 |
| 4 | 1.0 | 0.7 |

**Table 1 |** Initial stellar masses of four binary models

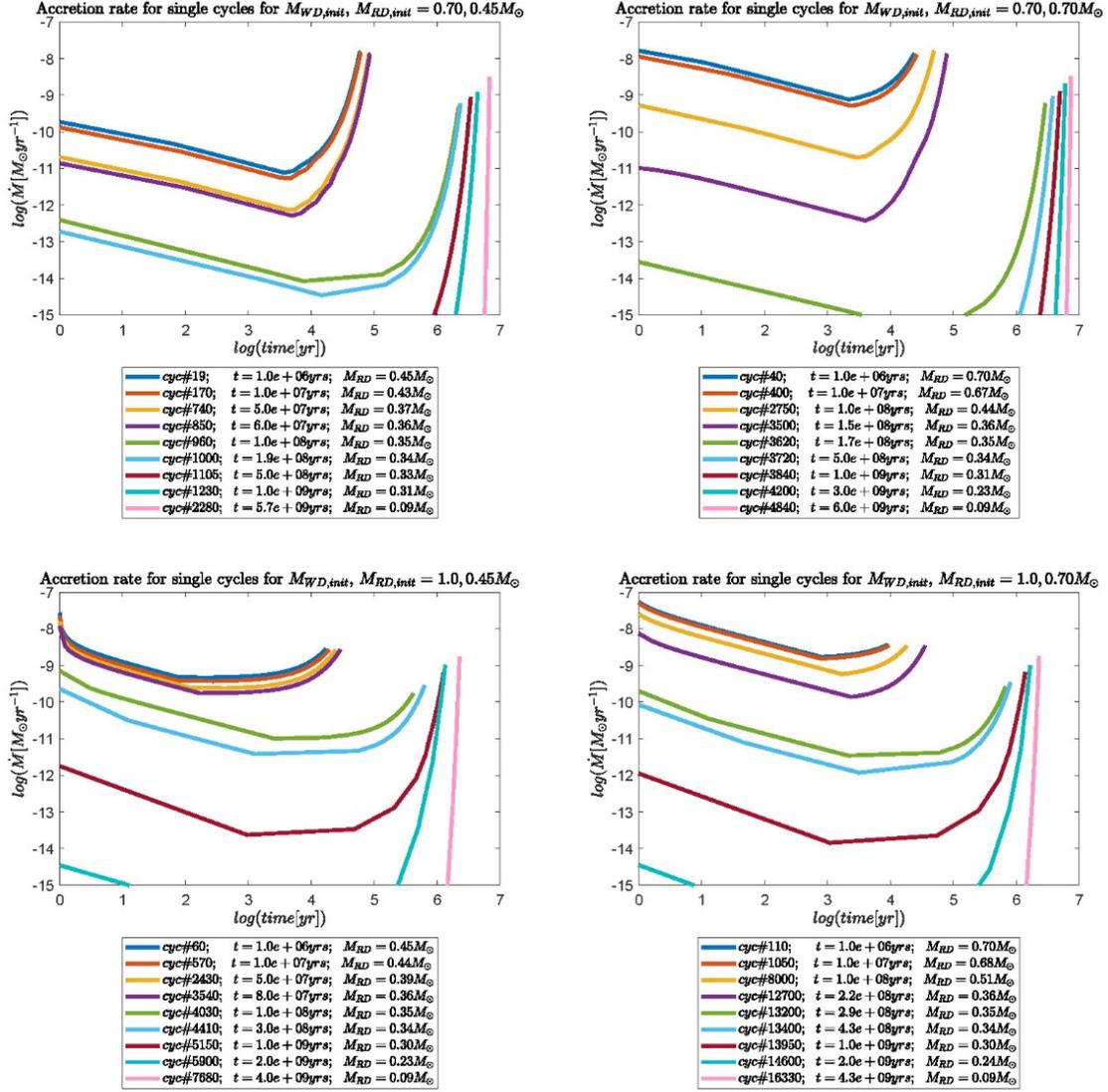

**Figure 1** | Cyclic variation of the accretion rate onto the white dwarf during nine individual nova cycles spanning the entire, multi-Gyr evolution for each of the four models listed in Table 1. The time is counted in each case from the beginning of the corresponding nova cycle. Both axes are on logarithmic scales.

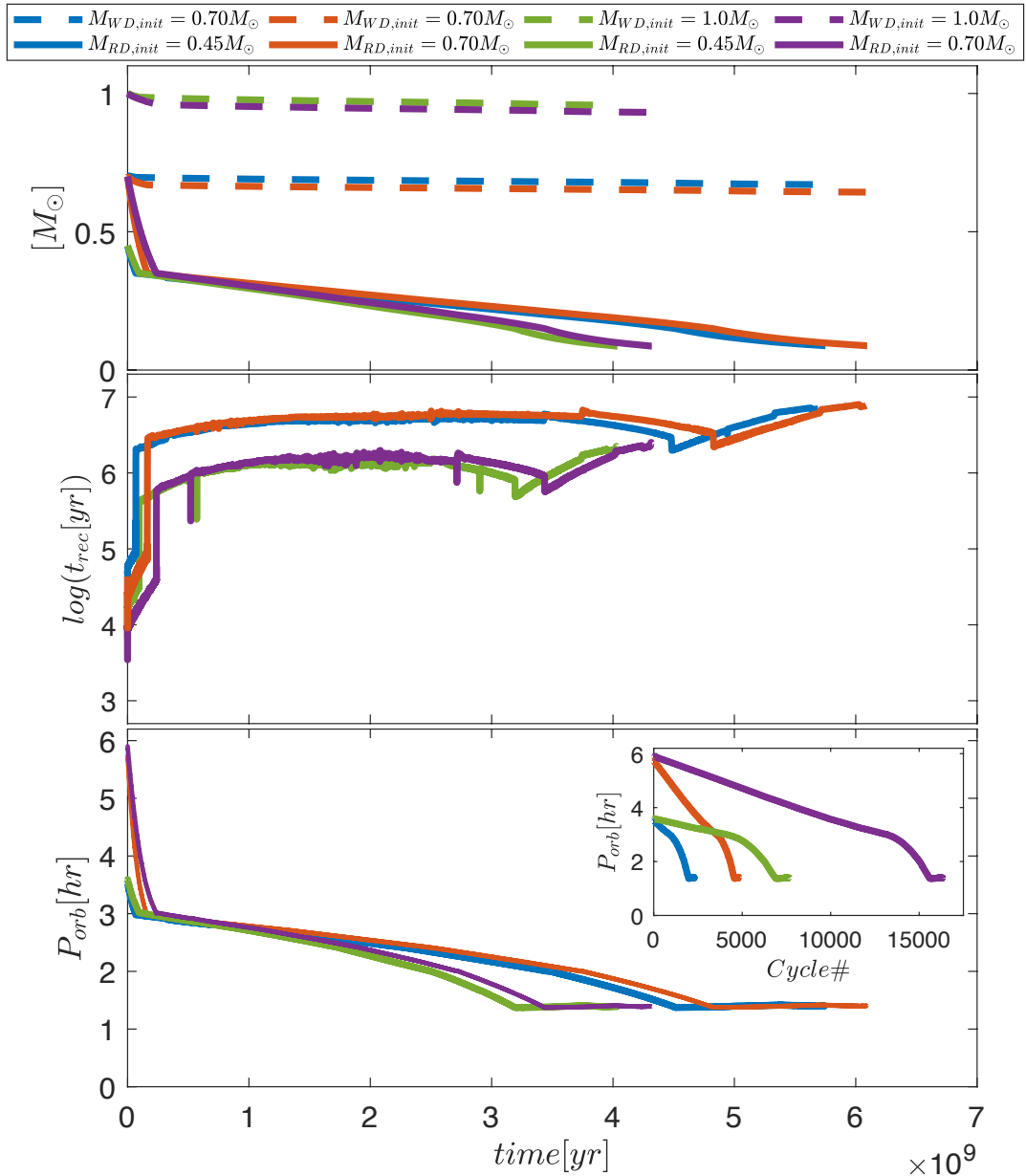

**Figure 2** | Top: Evolution of the binary masses - white dwarf (dashed lines) and red dwarf (solid lines) for the models of Table 1, color-coded as follows: model 1 - blue; model 2 - red; model 3 - green; model 4 - purple. White dwarf masses barely change over the entire evolution of these cataclysmic variables. Cataclysmic variables with initially similar white

dwarf masses but very different red dwarf masses converge to become nearly identical binaries as their red dwarf masses become similar. Center: The time between successive nova eruptions (recurrence time) versus time for each of the four models. Red dwarfs more massive than $\sim 0.35\ M_\odot$ transfer mass quickly enough that their recurrence times are of order $10^4$ yr. Once red dwarfs fall below this mass and become fully convective their average mass transfer rates drop by 2-3 orders of magnitude, causing their recurrence times to lengthen to $10^6 - 10^7$ yr. Bottom: Evolution of the binaries' orbital periods; the insert shows the orbital period vs. cycle number.

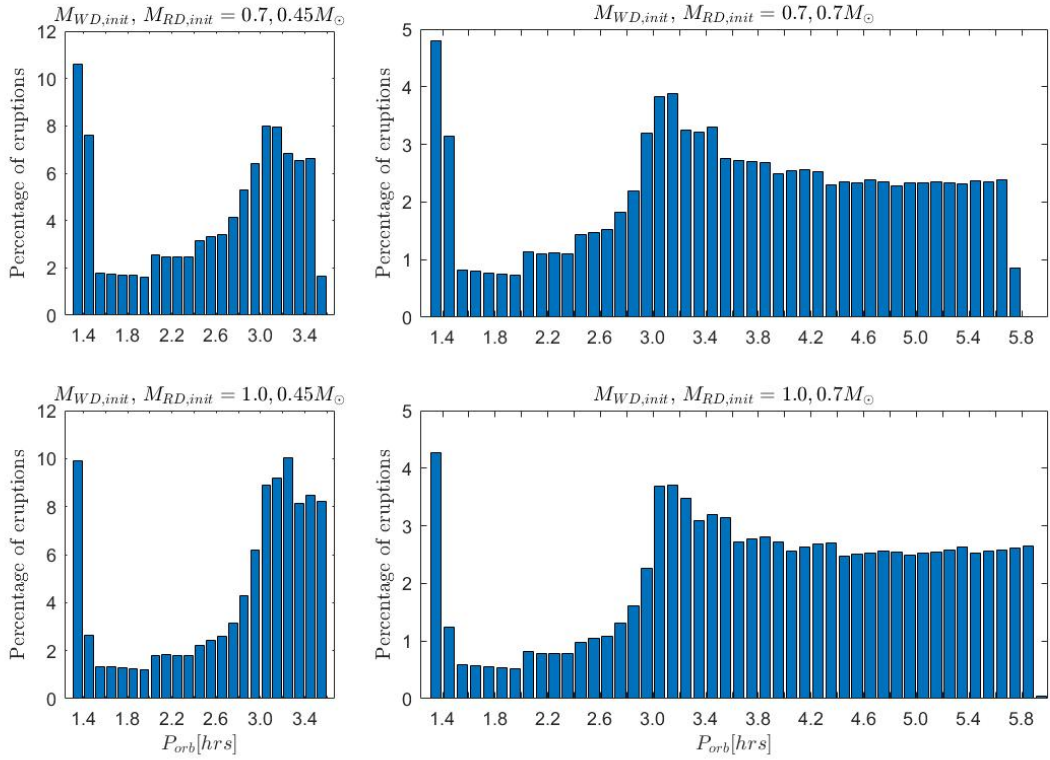

**Figure 3** | Percentage of nova eruptions for which a binary system is found in an orbital period interval, for each of the four models. Regardless of the initial white dwarf and red dwarf masses and separations, the large majority of nova eruptions occur when systems' orbital periods exceed 3 hours.

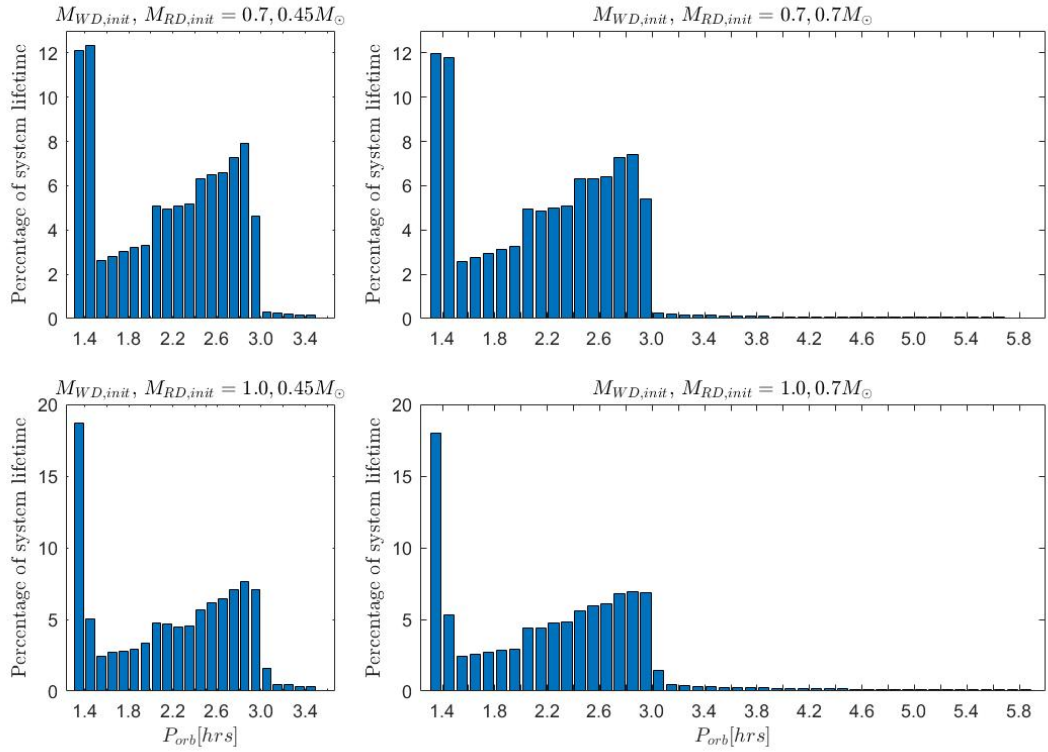

**Figure 4** Percentage of cataclysmic variable binary system lifetime spent at different orbital periods. All cataclysmic variables spend the vast majority of their lifetimes at orbital periods shorter than 3 hours, where the time between successive nova eruptions is 100 time longer than that for cataclysmic variables with periods longer than 3 hours.

**Methods**

**The computer codes and red dwarf models**

The self-consistent simulations of novae were carried out by means of two originally independent codes that have been modified and combined. The first, a hydrostatic stellar evolution code[26], was used to build a database of parameters of red dwarf donor stars with solar metallicity and masses $M_{RD}$ ranging from just under $M_{RD} = 0.10 M_\odot$ to $M_{RD} = 0.7 M_\odot$ at intervals of $0.05 M_\odot$. The code was run for each $M_{RD}$ separately, beginning from the pre-MS stage, while recording, for each time step throughout evolution, the stellar parameters that are necessary to calculate the combined binary evolution: the stellar mass ($M_{RD}$), the stellar radius ($R_{RD}$), the effective temperature ($T_{eff}$), the stellar luminosity ($L_{RD}$), the surface layer density ($\rho_{srf}$), the pressure scale height ($H_P$), the extent of convective zones, and the evolution time ($t$).

This database of red dwarf models was used by the second code — a hydrodynamic Lagrangian nova evolution code[27,28,29]. The original nova evolution code (like all others in the literature) did not depend on or use the properties of the donor star. Instead, it assumed a constant rate of accretion ($\dot{M}$) that was not recalculated during the binary's evolution. To self-consistently model the simultaneous evolution of both stars the nova code was adapted to allow for a changing $\dot{M}$ by recalculating the binary separation $a$ *between* nova cycles due to the change in the stars' masses that occurred in the previous cycle, and at *each time step* during the accretion phase, due to gravitational radiation and magnetic braking. The effect of enhanced irradiation on $\dot{M}$ during the nova eruptions was taken into consideration as well.

### Angular momentum, stellar separation and mass transfer rate calculations

During the accretion phase, at the end of each time step $\delta t$, the orbital angular momentum is:

$$(1) \quad J(t + \delta t) = J(t) + \dot{J}_{GR}\delta t + \dot{J}_{MB}\delta t \ .$$

The changes due to gravitational radiation ($\dot{J}_{GR}$) and due to magnetic braking ($\dot{J}_{MB}$) are[30]:

$$(2) \quad \dot{J}_{GR} = -\frac{32}{5c^5}\left(\frac{2\pi G}{P_{orb}}\right)^{\frac{7}{3}} \frac{(M_{RD}M_{WD})^2}{(M_{RD}+M_{WD})^{\frac{2}{3}}}$$

$$(3) \quad \dot{J}_{MB} = -1.06 \times 10^{20} M_{RD} R_{RD}^4 P_{orb}^{-3},$$

where the orbital period $P_{orb}$ (in seconds) is calculated from Kepler's law $P_{orb}^2 = 4\pi^2 a(t)^3/G(M_{WD} + M_{RD})$. The new separation is:

$$(4) \quad a(t + \delta t) = \frac{J^2(t+\delta t)\,(M_{RD}+M_{WD})}{G(M_{RD}\,M_{WD})^2},$$

and the Roche Lobe radius ($R_{RL}$) for $M_{RD}$ is then determined, using the prescription[31]:

$$(5) \quad \frac{R_{RL}}{a} = \frac{0.49 q^{\frac{2}{3}}}{0.6 q^{\frac{2}{3}} + \ln\left(1 + q^{\frac{1}{3}}\right)},$$

where $q = M_{RD}/M_{WD}$. Finally, the accretion rate ($\dot{M}$) is calculated by using $R_{RL}$, along with the radius of the secondary ($R_{RD}$), its surface density ($\rho_{srf}$), pressure scale height $H_P$ and its effective temperature ($T_{eff}$)[32]:

$$(6) \quad \dot{M} = 2\pi e^{-\frac{1}{2}} \left(\frac{R_g T_{eff}}{\mu}\right)^{\frac{3}{2}} \frac{R_{RL}^3 \rho_{srf}}{GM_{RD}} (1.23 + 0.5 \log q^{-1}) e^{\frac{R_{RD}-R_{RL}}{H_P}}.$$

Magnetic braking ceases when stars become fully convective; this occurs in our zero age MS star models at $M_{RD} = 0.35 M_\odot$. The donor mass is eroded throughout evolution, eventually becoming $< 0.35 M_\odot$. After this occurs only gravitational radiation and nova mass loss are able to remove angular momentum from the binary to decrease the components' separation.

When each nova eruption concludes, and just before accretion resumes, the white dwarf and red dwarf masses are updated and the resulting change in separation is implemented:

$$(7) \quad \Delta a = 2 \left( \frac{m_{ej} - m_{acc}}{M_{WD}} + \frac{m_{acc}}{M_{RD}} \right),$$

where $m_{acc}$ and $m_{ej}$ are the accreted and ejected masses during the cycle that has just ended.

**Effects of Irradiation**

The mass transfer rate immediately following an eruption is corrected[17] to take account of irradiation of the donor by the high luminosity of the white dwarf during a nova eruption and its early decline.

The irradiation luminosity[17] is

$$(8) \quad L_{irr} = \frac{1}{2} \left[ 1 - \sin \left( \arccos \frac{R_{RD}}{a} \right) \right] L_{Edd}(M_{WD}),$$

where $L_{Edd}(M_{WD})$ is the white dwarf Eddington luminosity, representing the peak luminosity of the eruption. As a result of irradiation, the temperature in the outer layers of the irradiated red dwarf rises to $\sim T_{irr}$, given by[17]

$$(9)\ T_{irr} = \left(\frac{L_{irr}}{4\pi\sigma R_{RD}^2}\right)^{\frac{1}{4}},$$

which is of the order of $5 \times 10^4$K. The consequent expansion fraction ε of the irradiated red dwarf radius $R_{RD,irr}$ is larger for smaller radii, and in the range[17] 1- 3 %.

The mass loss rate of the irradiated star immediately following an outburst is given by

$$(10)\ \dot{M}_{irr,0} \propto T_{irr}^{\frac{3}{2}} e^{\frac{R_{RD}-R_{RL}}{H_P}} f(R_{RD,irr}),$$

where $f$ is a parametrized function[17] of the RL overflow of the bloated donor star; it includes the effects of white dwarf masses different[17] than $1.25 M_\odot$ and it has relatively little effect on $\dot{M}_{irr,0}$. The major effect of irradiation is of the order of $(T_{irr}/T_{eff})^{\frac{3}{2}}$, which (from equations 6 and 10) is an enhancement factor of ~100.

Once the nova eruption starts declining, the irradiation from the white dwarf decreases by orders of magnitude and its effect on $\dot{M}$ decreases accordingly. The rate of decrease is[17]

$$(10)\quad \frac{\dot{M}_{irr}}{\dot{M}_{irr,0}} = \left(\frac{t}{t_0}\right)^{-0.45},$$

where $t_0$ is the period of time during which the star underwent enhanced irradiation, which is well approximated by the mass-loss time of the nova eruption[17]. This correction to $\dot{M}$ is implemented at each time step; hundreds to thousands of years after a nova eruption it becomes negligible.

A simulation begins with given initial stellar masses $M_{WD}$ and $M_{RD}$, separated by $a = R_{RL}$. Throughout the entire simulation of thousands of nova eruptions (and millions of timesteps for each binary) $\dot{M}$ evolves as described above, with no external intervention.

The simulations' orders of magnitude cyclical changes in $\dot{M}$ are the direct result of the self-consistent and co-dependent, feedback-driven evolution of each star.

**Physical Effects not included in simulations**

Several mechanisms have been suggested as additional sinks of angular momentum from cataclysmic binaries: frictional angular momentum loss[14,33,34,35], consequential angular momentum loss[36], magnetically coupled angular momentum loss[37], and asymmetrical jets[37]. Frictional angular momentum loss is almost always negligible in size[33,34,35], while consequential angular momentum loss is invoked ad hoc[36] to reconcile standard gravitational radiation and magnetic braking angular momentum loss rate prescriptions with the paths of red dwarfs in mass-radius relations. Magnetically coupled angular momentum loss[37], and asymmetrical jets[37] were invoked to explain claimed (and very unexpected) orbital period decreases across nova eruptions. The models presented here form a baseline study incorporating only those mechanisms that are certain to be operative and important, and for which well-defined and quantitative theories exist. Thus none of the four extra angular momentum loss mechanisms noted above are included in the simulations of this paper.

**References**

26. Kovetz, A., Yaron, O. & Prialnik, D. A New, Efficient Stellar Evolution Code for Calculating Complete Evolutionary Tracks. *Mon. Not. Roy. Astron. Soc.* **395**, 1857–1874 (2009).

27. Prialnik, D. & Kovetz, A. An Extended Grid of Multicycle Nova Evolution Models. *The Astrophysical Journal* **445**, 789–810 (1995).

28. Epelstain, N., Yaron, O., Kovetz, A. & Prialnik, D. A Thousand and One Nova Outbursts. *Mon. Not. Roy. Astron. Soc.* **374**, 1449–1456 (2007).

29. Hillman, Y., Prialnik, D., Kovetz, A. & Shara, M. M. Observational Signatures of SNIa Progenitors, as Predicted by Models. *Mon. Not. Roy. Astron. Soc.* **446**, 1924–1930 (2015).

30. Paxton, B. et al. Modules for Experiments in Stellar Astrophysics (MESA): Binaries, Pulsations, and Explosions. *Astrophysical Journal Supplement* **220** 15-59 (2015).

31. Eggleton, P. Approximations to the Radii of Roche lobes. *The Astrophysical Journal* **268**, 368–369 (1983).

32. Ritter, H. Turning on and off mass transfer in cataclysmic binaries. *A&A* **202**, 93–100 (1988).

33. MacDonald, J. Postthermonuclear runaway angular momentum loss in cataclysmic binaries. *The Astrophysical Journal* **305**, 251-260 (1986).

34. Schenker, K., Kolb, U. and Ritter, H. Properties of discontinuous and nova-amplified mass transfer in cataclysmic binaries. *Mon. Not. Roy. Astron. Soc.* **297**, 633-647 (1998).

35. Liu, W.-M. and Li, X.-D. Can the friction of the nova envelope account for the extra angular momentum loss in cataclysmic variables? *The Astrophysical Journal* **870**, 22-30 (2019).

36. Schreiber, M.R., Zorotovic, M., and Wijnen, T.P.G. Three in one go: consequential angular momentum loss can solve major problems of CV evolution. *Mon. Not. Roy. Astron. Soc.* **455**, L16-L20 (2016).

37. Schaefer, B. et al. Precise measures of orbital period, before and after nova eruption for QZ Aur. *Mon. Not. Roy. Astron. Soc.* **487**, 1120-1139 (2019).